\documentclass[10pt,journal,compsoc]{IEEEtran}

\usepackage[colorlinks,urlcolor=blue,linkcolor=blue,citecolor=blue]{hyperref}
\usepackage{color}
\newcommand{\review}[1]{{\color{black}#1\normalfont}}
\newcommand{\minorreview}[1]{{\color{black}#1\normalfont}}

\usepackage[table,xcdraw]{xcolor}
\usepackage{listings}
\usepackage{tikz}

\usepackage{booktabs}   
\usepackage{subcaption} 

\usepackage[labelfont=bf]{caption}
\usepackage[utf8]{inputenc} 
\usepackage[T1]{fontenc}
\usepackage{microtype}

\usepackage{enumitem}
\usepackage{color}
\usepackage{amssymb}
\usepackage{multirow}
\usepackage{etoolbox}
\usepackage[plain]{algorithm}
\usepackage[noend]{algorithmic}
\usepackage{syntax}
\usepackage{comment}
\usepackage{amsthm}
\usepackage{amsmath}
\usepackage{xspace}
\usepackage{tcolorbox}
\usepackage[all]{nowidow}
\usepackage{booktabs}
\usepackage{algorithmic}
\usepackage[english]{babel}
\usepackage{hyphenat}
\usepackage{ulem}
\usepackage{listings}
\usepackage{mdframed}

\definecolor{pblue}{rgb}{0.13,0.13,1}
\definecolor{pgreen}{rgb}{0,0.5,0}
\definecolor{pred}{rgb}{0.9,0,0}
\definecolor{pgrey}{rgb}{0.46,0.45,0.48}
\definecolor{darkblue}{rgb}{0.0, 0.0, 0.55}
\definecolor{light-gray}{gray}{0.9}

\newcommand{\lstbg}[3][0pt]{{\fboxsep#1\colorbox{#2}{\strut #3}}}

\lstdefinelanguage{diff}{
  basicstyle=\ttfamily\small,
  morecomment=[f][\lstbg{pred!20}]-,
  morecomment=[f][\lstbg{pgreen!20}]+,
  morecomment=[f][\textit]{@@},
  morecomment=[f][\textit]{---},
  morecomment=[f][\textit]{+++},
}

\newcommand{\gemma}{Gemma-2-9B}
\newcommand{\llama}{Llama-3.2-3B}
\newcommand{\phimodel}{Phi-4-14B}
\newcommand{\deepseek}{DeepSeek-R1-14B}

\newcommand{\othree}{o3}
\newcommand{\claude}{Claude-4-Sonnet}
\newcommand{\gemini}{Gemini-2.5-Pro}

\newcommand{\totalExamples}{150}
\newcommand{\totalTypes}{5}
\newcommand{\totalRefactor}{70.6\%{}} 

\newcommand{\othreeRefactor}{74.7\%{}} 
\newcommand{\claudeRefactor}{59.3\%{}} 
\newcommand{\geminiRefactor}{72.0\%{}} 

\newcommand{\gemmaDetect}{98\%{}} 
\newcommand{\gemmaRefactor}{40.7\%{}} 

\newcommand{\llamaRefactor}{22\%{}} 

\newcommand{\phiDetect}{96\%{}} 
\newcommand{\phiDetectPassFive}{96\%{}} 
\newcommand{\phiRefactor}{55.3\%{}} 
\newcommand{\phiRefactorPassFive}{75.3\%{}}

\newcommand{\deepseekDetect}{78\%{}} 
\newcommand{\deepseekRefactor}{39.3\%{}} 

\begin{document}

\title{Agentic \review{LMs}: Hunting Down Test Smells}

\author{
    \IEEEauthorblockN{
        Rian Melo\IEEEauthorrefmark{3}, Pedro Sim\~{o}es\IEEEauthorrefmark{3}, Rohit Gheyi\IEEEauthorrefmark{3},Marcelo d'Amorim\IEEEauthorrefmark{6},Márcio Ribeiro\IEEEauthorrefmark{2},Gustavo Soares\IEEEauthorrefmark{4},Eduardo Almeida\IEEEauthorrefmark{5},Elvys Soares\IEEEauthorrefmark{1}\\
    }
    \IEEEauthorblockA{
        \IEEEauthorrefmark{3}UFCG, Brazil\\Email: \{rian.melo,pedro.henrique.lima.simoes\}@ccc.ufcg.edu.br and rohit@dsc.ufcg.edu.br\\
    }
    \IEEEauthorblockA{
        \IEEEauthorrefmark{6}NCSU, USA\\Email: mdamori@ncsu.edu\\          
    }
    \IEEEauthorblockA{
        \IEEEauthorrefmark{2}UFAL, Brazil\\Email: marcio@ic.ufal.br\\
    }
    \IEEEauthorblockA{
        \IEEEauthorrefmark{4}Microsoft, USA\\Email: gsoares@microsoft.com\\
    }
    \IEEEauthorblockA{
        \IEEEauthorrefmark{5}UFBA, Brazil\\Email: eduardo.almeida@ufba.br\\   
    }
    \IEEEauthorblockA{
        \IEEEauthorrefmark{1}IFAL, Brazil\\Email: elvys.soares@ifal.edu.br\\
    }    
}

\maketitle
\begin{abstract}

Test smells reduce test suite reliability and complicate maintenance.  
While many methods detect test smells, few support automated removal, and most rely on static analysis or machine learning.  
This study evaluates \review{models with relatively small parameter counts}—\llama{}, \gemma{}, \deepseek{}, and \phimodel{}—for their ability to detect and refactor test smells using agent-based workflows.
We assess workflows with one, two, and four agents over \totalExamples{} instances of \totalTypes{} common smells from real-world Java projects. Our approach generalizes to Python, Golang\review{, and JavaScript}.  
All models detected nearly all instances, with \phimodel{} achieving the best refactoring accuracy (pass@5 of \phiRefactorPassFive{}). \phimodel{} with four-agents performed within 5\% of proprietary \review{LLMs} (single-agent). 
Multi-agent setups outperformed single-agent ones in three of five smell types, though for \textit{Assertion Roulette}, one agent sufficed.  
We submitted pull requests with \phimodel{}-generated code to open-source projects and \review{six} were merged.

\end{abstract}


\maketitle

\section{Introduction}
\label{sec:introduction}

Test smells are design flaws in test code that compromise reliability and hinder maintenance~\cite{van2001refactoring}. They can reduce the effectiveness of test suites. These issues are prevalent in both open-source and industrial settings~\cite{Peruma2019Distribution}.
Common test smells include undocumented assertions (\textit{Assertion Roulette}), complex conditional logic (\textit{Conditional Test Logic}), duplicated assertions (\textit{Duplicate Assert}), and hard-coded values (\textit{Magic Number})~\cite{Peruma2019Distribution}. 

Existing tools for detecting test smells typically rely on static analysis~\cite{aljedaani2021test} or machine learning~\cite{pontillo2024,aljedaani2021test}, but these approaches are often hard to adapt to new smells or other languages~\cite{aljedaani2021test}. Recent advances in foundation models have opened new possibilities in software engineering, yet their use in test smell refactoring -- especially through collaborative agentic workflows -- remains underexplored.

In this work, we evaluate open models (\llama{}, \gemma{}, \deepseek{}, and \phimodel{}) for automatically detecting and refactoring test smells through agentic workflows. 
Our method supports one, two, or four agents and is easily extensible: new smells and refactorings can be defined in natural language. We evaluate \totalExamples{} instances of \totalTypes{} frequent test smells from real-world Java projects (Section~\ref{sec:methodology}).

Results show that all models detected nearly all test smell instances (pass@5 of \phiDetect{} with four-agents), with \phimodel{} achieving the highest refactoring accuracy (pass@5 of \phiRefactorPassFive{}). This performance is within 5\% of proprietary \review{LLMs} using a single-agent. Multi-agent setups outperformed single-agent ones in three of five smell types, though for \textit{Assertion Roulette}, a single-agent configuration proved more effective.

Preliminary results demonstrate promising generalization across programming languages -- specifically Python, Golang\review{, and JavaScript} -- using the same setup. Beyond automating the detection and refactoring of test smells, our approach also enables developers to interact with models to better understand the rationale behind suggested changes.
All artifacts are publicly available online.\footnote{\label{fn:exemplo}\url{ https://zenodo.org/records/17285750}}

\section{Test Smells}
\label{sec:test-smells}

\review{
Test smells are recurring patterns that reduce test clarity and maintainability.
\textit{Assertion Roulette}~\cite{van2001refactoring} occurs when a test method contains more than one assertion statement without an explanation or message (parameter in the assertion method). To mitigate this smell, developers should add a message to each assertion.
\textit{Conditional Test Logic}~\cite{mesaros} arises when a test method contains one or more control statements (i.e., conditional expression, and loop statements). This can be addressed by having the developer create a dedicated test method for each condition.
\textit{Duplicate Assert}~\cite{Peruma2019Distribution} appears when a test method contains more than one assertion statement with the same parameters. To address this, developers should split assertions that test different scenarios or states into separate tests for clarity.
\textit{Exception Handling} occurs when a test method contains either a \texttt{throw} statement or at least a \texttt{catch} clause. To avoid this smell, use the testing framework's features (e.g., \texttt{assertThrows}) instead of manually catching or throwing exceptions.
\textit{Magic Number} occurs when a test method contains an assertion with a numeric literal as an argument. Refactoring involves extracting and initializing all magic numbers into constants or local variables with descriptive names.
}

\section{Methodology}
\label{sec:methodology}

The primary goal of this study is to evaluate the effectiveness of an agentic approach using \review{models} to detect and remove test smells. We analyzed test methods from 11 real-world open-source GitHub projects using JUnit-5 previously studied by Soares et al.~\cite{SoaresRGAS23}, including \textit{janusgraph}, \textit{quarkus}, \textit{testcontainers-java}, \textit{opengrok}, \textit{jenkins}, \textit{lettuce}, \textit{Mindustry}, \textit{data-transfer-project}, \textit{Activiti}, \textit{flowable-engine}, and \textit{skywalking}. \review{We limited our analysis to tests with up to 30 LOC, as 89\% of test cases in the studied projects fall within this range.} We focused on \review{\totalTypes{} common test smell types~\cite{Peruma2019Distribution}}, selecting 30 real-world examples for each.
\review{We evaluate smaller models up to 14B parameters:} \llama{}, \gemma{}, \deepseek{}, and \phimodel{} using their default configurations. \review{All models were accessed via the Ollama platform and executed locally on a MacBook Pro M3 with 18GB of RAM (July 2025).}

\begin{figure*}[htbp]
  \centering
  \includegraphics[scale=0.06]{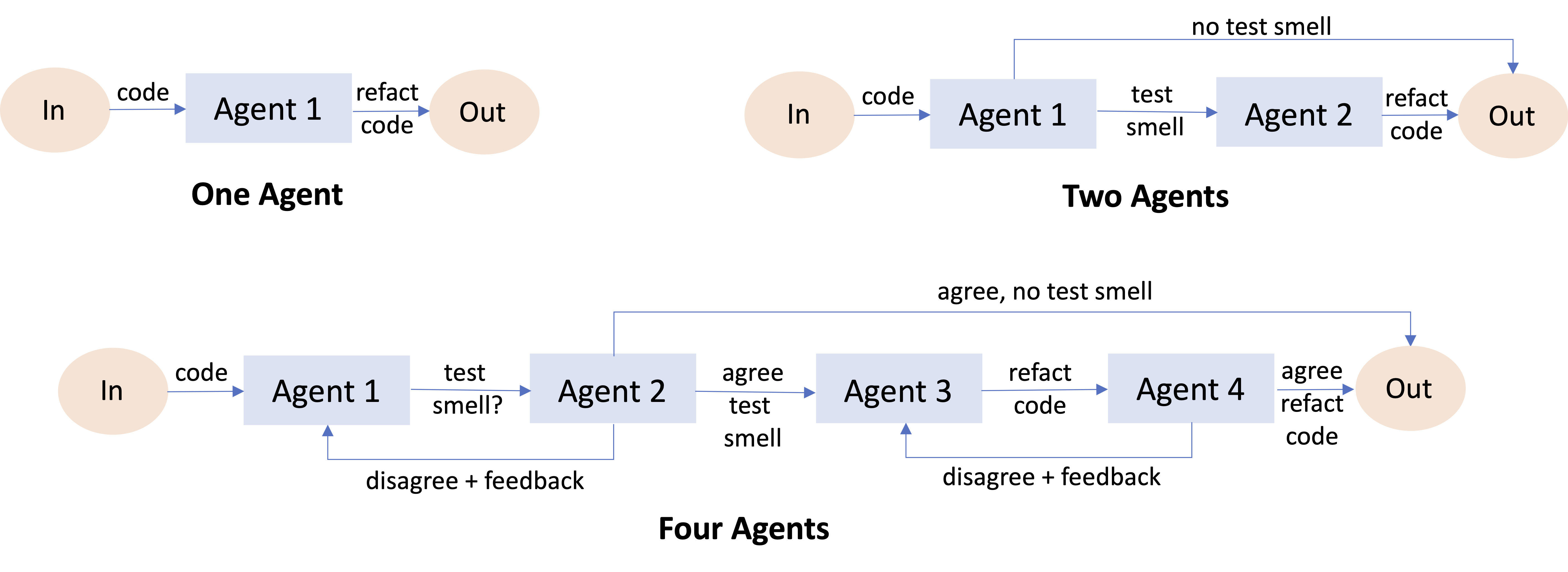}
  \caption{Agentic workflow with one, two and four agents to detect and remove test smells.}
  \label{fig:agentic4}  
\end{figure*}

Agent communication is managed using the LangChain API, and we apply prompting techniques such as Role (Persona) and Chain-of-Thought~\cite{prompt-techniques2} to enhance reasoning and contextual understanding.
The four-agent configuration (Figure~\ref{fig:agentic4}) distributes responsibilities as follows: Agent$_{1}$ detects potential smells; Agent$_{2}$ confirms the detection; Agent$_{3}$ performs the refactoring; and \review{Agent$_{4}$ verifies the correctness of the refactored code.} If there is disagreement, agents engage in an Evaluator-Optimizer loop\footnote{\url{https://www.anthropic.com/research/building-effective-agents}}, repeated up to three times.

Agent$_{1}$ utilizes the following prompt to detect test smells.
\review{The prompts use \texttt{variables} to represent relevant information required for each analysis. For instance, \texttt{test_smell_definition_and_refactoring} comprises two sentences: one that defines the smell and another that suggests an appropriate refactoring (see Section~\ref{sec:test-smells} for example sentences).
}
\begin{mdframed}[backgroundcolor=cyan!5, linecolor=black, linewidth=0.5pt]
\footnotesize
\noindent You are a coding assistant with many years of experience that detects test smells. \\
\review{\texttt{test_smell_definition_and_refactoring}} \\
Your goal is to determine if the provided test code exhibits the test smell ``\review{\texttt{test_smell_name}}''. \\
\review{\texttt{code}} \\
\noindent Next I may give you further details. \\
\review{\texttt{agent_feedback}} \\
\noindent If the test code contains \review{\texttt{test_smell_name}}, respond with EXACTLY ``YES'' on the first line and explain why. Ignore code comments. If it does not contain, say EXACTLY ``NO'' on the first line and explain why not.
\end{mdframed}

The prompt used by Agent$_{2}$ is designed to evaluate the output generated by Agent$_{1}$.

\begin{mdframed}[backgroundcolor=cyan!5, linecolor=black, linewidth=0.5pt]
\footnotesize
\noindent You are a coding expert reviewing the detection of a test smell. Consider the following test smell. \\
\review{\texttt{test_smell_definition_and_refactoring}} \\
\noindent A previous agent analyzed the following test code. \\
\review{\texttt{code}} \\
\noindent It gave the following answer. \\
\review{\texttt{agent_answer}} \\
\noindent Your goal is to evaluate if the previous detection by another agent is correct and justified. Ignore code comments. If you do not agree, answer NO and explain what's wrong with it and what to correct. If yes, just say YES.
\end{mdframed}

To remove a test smell, Agent$_{3}$ utilizes the prompt outlined next. During the second or third iteration, it incorporates feedback from Agent$_{4}$. 
\vspace{0.1cm}
\begin{mdframed}[backgroundcolor=cyan!5, linecolor=black, linewidth=0.5pt]
\footnotesize
\noindent You are a coding assistant specializing in test code analysis and refactoring, with many years of experience. \\
\review{\texttt{test_smell_definition_and_refactoring}} \\
Your task is as follows.
First analyze the provided test code to resolve test smell occurrences ``\review{\texttt{test_smell_name}}''. If there is no smell, output the original code unchanged. Second ensure the test preserves the same behavior, but is free of \review{\texttt{test_smell_name}}. Third output only the final refactored code, valid under JUnit 5. Finally check the refactored version does not introduce compilation errors. \\
\noindent Provide only the final refactored code, with no additional explanation or text. \\
\noindent Code to analyze: \\
\review{\texttt{code}} \\
Next I may provide you further details. \\
\review{\texttt{agent_feedback}}
\end{mdframed}

To evaluate the \review{\texttt{refactored_code}} proposed by Agent$_{3}$, Agent$_{4}$ utilizes the following prompt. 

\begin{mdframed}[backgroundcolor=cyan!5, linecolor=black, linewidth=0.5pt]
\footnotesize
\noindent You are a code reviewer specializing in JUnit 5 test smells. \\
\review{\texttt{test_smell_definition_and_refactoring}} \\
\noindent Analyze the following code. \\
\review{\texttt{refactored_code}} \\
\noindent Your task is to check three conditions. First check the code does not have the test smell \review{\texttt{test_smell_name}}. Second verify the code follows JUnit 5 specification. Finally confirms the code does not have compilation errors. If the code satisfy all conditions, respond with EXACTLY ``YES'' on the first line. If not, respond with EXACTLY ``NO'' on the first line, then explain in one or two sentences why. Let's think step by step.
\end{mdframed}

The two-agent setup follows the same workflow but omits Agents$_{2,4}$. Agent$_{1}$ handles detection, and Agent$_{3}$ performs the refactoring, without validation or feedback loops.
In the single-agent setup, Agent$_{1}$ analyzes the test code, identifies potential smells, and applies the corresponding refactoring.

\section{Evaluation}
\label{sec:evaluation}

Table~\ref{tab:summary} summarizes the detailed performance \review{(pass@1)} of the single-agent and multi-agent approaches for detecting and refactoring test smells. \minorreview{In our setting, pass@1 measures whether the model’s first attempt correctly detects or refactors a test smell, with correctness independently verified by two authors and cross-checked by a third.}

\begin{table*}[]
\caption{Detection and refactoring performance \review{(pass@1)} of test smells by one, two, and four agents using \llama{} (Lla), \gemma{} (Gem), \deepseek{} (DS) and \phimodel{} (Phi).
}
\label{tab:summary}

\resizebox{\textwidth}{!}{ 

\begin{tabular}{cr|rrrr|rrrrrrrr|rrrrrrrr|}
\cline{3-22}
\arrayrulecolor{white}
\rowcolor[HTML]{000000} 
\multicolumn{1}{l}{\cellcolor[HTML]{FFFFFF}}                                             & \multicolumn{1}{l|}{\cellcolor[HTML]{FFFFFF}}                                         & \multicolumn{4}{c|}{\cellcolor[HTML]{000000}{\color[HTML]{FFFFFF} \textbf{One Agent}}}                                                                                                                                                                                                                                                             & \multicolumn{8}{c|}{\cellcolor[HTML]{000000}{\color[HTML]{FFFFFF} \textbf{Two Agents}}}                                                                                                                                                                                                                                                                                                                                                                                                                                                                                                                                                                                                                 & \multicolumn{8}{c|}{\cellcolor[HTML]{000000}{\color[HTML]{FFFFFF} \textbf{Four Agents}}}                                                                                                                                                                                                                                                                                                                                                                                                                                                                                                                                                                                                                \\ \cline{3-22} 
\rowcolor[HTML]{000000} 
\multicolumn{1}{l}{\cellcolor[HTML]{FFFFFF}}                                             & \multicolumn{1}{l|}{\cellcolor[HTML]{FFFFFF}}                                         & \multicolumn{4}{c|}{\cellcolor[HTML]{000000}{\color[HTML]{FFFFFF} \textbf{Detect and Refactoring}}}                                                                                                                                                                                                                                                & \multicolumn{4}{c|}{\cellcolor[HTML]{000000}{\color[HTML]{FFFFFF} \textbf{Detect}}}                                                                                                                                                                                                                                                                & \multicolumn{4}{c|}{\cellcolor[HTML]{000000}{\color[HTML]{FFFFFF} \textbf{Refactoring}}}                                                                                                                                                                                                                                                           & \multicolumn{4}{c|}{\cellcolor[HTML]{000000}{\color[HTML]{FFFFFF} \textbf{Detect}}}                                                                                                                                                                                                                                                                & \multicolumn{4}{c|}{\cellcolor[HTML]{000000}{\color[HTML]{FFFFFF} \textbf{Refactoring}}}                                                                                                                                                                                                                                                           \\ \hline
\rowcolor[HTML]{000000} 
\multicolumn{1}{|c|}{\cellcolor[HTML]{000000}{\color[HTML]{FFFFFF} \textbf{\review{Test smell}}}} & \multicolumn{1}{c|}{\cellcolor[HTML]{000000}{\color[HTML]{FFFFFF} \textbf{Sub.}}} & \multicolumn{1}{c|}{\cellcolor[HTML]{000000}{\color[HTML]{FFFFFF} \textbf{Phi}}} & \multicolumn{1}{c|}{\cellcolor[HTML]{000000}{\color[HTML]{FFFFFF} \textbf{DS}}} & \multicolumn{1}{c|}{\cellcolor[HTML]{000000}{\color[HTML]{FFFFFF} \textbf{Gem}}} & \multicolumn{1}{c|}{\cellcolor[HTML]{000000}{\color[HTML]{FFFFFF} \textbf{Lla}}} & \multicolumn{1}{c|}{\cellcolor[HTML]{000000}{\color[HTML]{FFFFFF} \textbf{Phi}}} & \multicolumn{1}{c|}{\cellcolor[HTML]{000000}{\color[HTML]{FFFFFF} \textbf{DS}}} & \multicolumn{1}{c|}{\cellcolor[HTML]{000000}{\color[HTML]{FFFFFF} \textbf{Gem}}} & \multicolumn{1}{c|}{\cellcolor[HTML]{000000}{\color[HTML]{FFFFFF} \textbf{Lla}}} & \multicolumn{1}{c|}{\cellcolor[HTML]{000000}{\color[HTML]{FFFFFF} \textbf{Phi}}} & \multicolumn{1}{c|}{\cellcolor[HTML]{000000}{\color[HTML]{FFFFFF} \textbf{DS}}} & \multicolumn{1}{c|}{\cellcolor[HTML]{000000}{\color[HTML]{FFFFFF} \textbf{Gem}}} & \multicolumn{1}{c|}{\cellcolor[HTML]{000000}{\color[HTML]{FFFFFF} \textbf{Lla}}} & \multicolumn{1}{c|}{\cellcolor[HTML]{000000}{\color[HTML]{FFFFFF} \textbf{Phi}}} & \multicolumn{1}{c|}{\cellcolor[HTML]{000000}{\color[HTML]{FFFFFF} \textbf{DS}}} & \multicolumn{1}{c|}{\cellcolor[HTML]{000000}{\color[HTML]{FFFFFF} \textbf{Gem}}} & \multicolumn{1}{c|}{\cellcolor[HTML]{000000}{\color[HTML]{FFFFFF} \textbf{Lla}}} & \multicolumn{1}{c|}{\cellcolor[HTML]{000000}{\color[HTML]{FFFFFF} \textbf{Phi}}} & \multicolumn{1}{c|}{\cellcolor[HTML]{000000}{\color[HTML]{FFFFFF} \textbf{DS}}} & \multicolumn{1}{c|}{\cellcolor[HTML]{000000}{\color[HTML]{FFFFFF} \textbf{Gem}}} & \multicolumn{1}{c|}{\cellcolor[HTML]{000000}{\color[HTML]{FFFFFF} \textbf{Lla}}} \\ \hline
\arrayrulecolor{black}
\multicolumn{1}{|c|}{Assertion Roulette}                                                 & 30                                                                                    & \multicolumn{1}{r|}{93.3\%}                                                      & \multicolumn{1}{r|}{56.7\%}                                                           & \multicolumn{1}{r|}{66.7\%}                                                        & 20.0\%                                                                             & \multicolumn{1}{r|}{100.0\%}                                                     & \multicolumn{1}{r|}{86.7\%}                                                           & \multicolumn{1}{r|}{96.7\%}                                                        & \multicolumn{1}{r|}{100.0\%}                                                       & \multicolumn{1}{r|}{70.0\%}                                                      & \multicolumn{1}{r|}{33.3\%}                                                           & \multicolumn{1}{r|}{36.7\%}                                                        & 26.7\%                                                                             & \multicolumn{1}{r|}{100.0\%}                                                     & \multicolumn{1}{r|}{86.7\%}                                                           & \multicolumn{1}{r|}{96.7\%}                                                        & \multicolumn{1}{r|}{100.0\%}                                                       & \multicolumn{1}{r|}{73.3\%}                                                      & \multicolumn{1}{r|}{33.3\%}                                                           & \multicolumn{1}{r|}{36.7\%}                                                        & 26.7\%                                                                             \\ \hline
\multicolumn{1}{|c|}{Cond. Test Logic}                                             & 30                                                                                    & \multicolumn{1}{r|}{3.3\%}                                                       & \multicolumn{1}{r|}{6.7\%}                                                            & \multicolumn{1}{r|}{6.7\%}                                                         & 0.0\%                                                                              & \multicolumn{1}{r|}{96.7\%}                                                      & \multicolumn{1}{r|}{93.3\%}                                                           & \multicolumn{1}{r|}{100.0\%}                                                       & \multicolumn{1}{r|}{100.0\%}                                                       & \multicolumn{1}{r|}{13.3\%}                                                      & \multicolumn{1}{r|}{0.0\%}                                                            & \multicolumn{1}{r|}{13.3\%}                                                        & 0.0\%                                                                              & \multicolumn{1}{r|}{96.7\%}                                                      & \multicolumn{1}{r|}{93.3\%}                                                           & \multicolumn{1}{r|}{100.0\%}                                                       & \multicolumn{1}{r|}{100.0\%}                                                       & \multicolumn{1}{r|}{\review{10.0}\%}                                                        & \multicolumn{1}{r|}{3.3\%}                                                            & \multicolumn{1}{r|}{16.7\%}                                                        & 3.3\%                                                                              \\ \hline
\multicolumn{1}{|c|}{Duplicate Assert}                                                   & 30                                                                                    & \multicolumn{1}{r|}{40.0\%}                                                      & \multicolumn{1}{r|}{36.7\%}                                                           & \multicolumn{1}{r|}{23.3\%}                                                        & 10.0\%                                                                             & \multicolumn{1}{r|}{100.0\%}                                                     & \multicolumn{1}{r|}{50.0\%}                                                           & \multicolumn{1}{r|}{100.0\%}                                                       & \multicolumn{1}{r|}{100.0\%}                                                       & \multicolumn{1}{r|}{70.0\%}                                                      & \multicolumn{1}{r|}{20.0\%}                                                           & \multicolumn{1}{r|}{56.7\%}                                                        & 10.0\%                                                                             & \multicolumn{1}{r|}{93.3\%}                                                      & \multicolumn{1}{r|}{60.0\%}                                                           & \multicolumn{1}{r|}{100.0\%}                                                       & \multicolumn{1}{r|}{100.0\%}                                                       & \multicolumn{1}{r|}{60.0\%}                                                      & \multicolumn{1}{r|}{20.0\%}                                                           & \multicolumn{1}{r|}{60.0\%}                                                        & 10.0\%                                                                             \\ \hline
\multicolumn{1}{|c|}{Exception Handling}                                                 & 30                                                                                    & \multicolumn{1}{r|}{43.3\%}                                                      & \multicolumn{1}{r|}{13.3\%}                                                           & \multicolumn{1}{r|}{\review{40.0}\%}                                                          & 36.7\%                                                                             & \multicolumn{1}{r|}{90.0\%}                                                      & \multicolumn{1}{r|}{50.0\%}                                                           & \multicolumn{1}{r|}{93.3\%}                                                        & \multicolumn{1}{r|}{100.0\%}                                                       & \multicolumn{1}{r|}{43.3\%}                                                      & \multicolumn{1}{r|}{23.3\%}                                                           & \multicolumn{1}{r|}{43.3\%}                                                        & 30.0\%                                                                             & \multicolumn{1}{r|}{86.7\%}                                                      & \multicolumn{1}{r|}{50.0\%}                                                           & \multicolumn{1}{r|}{90.0\%}                                                        & \multicolumn{1}{r|}{100.0\%}                                                       & \multicolumn{1}{r|}{40.0\%}                                                      & \multicolumn{1}{r|}{23.3\%}                                                           & \multicolumn{1}{r|}{40.0\%}                                                        & 36.7\%                                                                             \\ \hline
\multicolumn{1}{|c|}{Magic Number}                                                       & 30                                                                                    & \multicolumn{1}{r|}{73.3\%}                                                      & \multicolumn{1}{r|}{83.3\%}                                                           & \multicolumn{1}{r|}{\review{30.0}\%}                                                          & 33.3\%                                                                             & \multicolumn{1}{r|}{93.3\%}                                                      & \multicolumn{1}{r|}{100.0\%}                                                          & \multicolumn{1}{r|}{100.0\%}                                                       & \multicolumn{1}{r|}{100.0\%}                                                       & \multicolumn{1}{r|}{80.0\%}                                                      & \multicolumn{1}{r|}{90.0\%}                                                           & \multicolumn{1}{r|}{43.3\%}                                                        & 40.0\%                                                                             & \multicolumn{1}{r|}{90.0\%}                                                      & \multicolumn{1}{r|}{100.0\%}                                                          & \multicolumn{1}{r|}{100.0\%}                                                       & \multicolumn{1}{r|}{100.0\%}                                                       & \multicolumn{1}{r|}{70.0\%}                                                      & \multicolumn{1}{r|}{90.0\%}                                                           & \multicolumn{1}{r|}{50.0\%}                                                        & 33.3\%                                                                             \\ \hline
\end{tabular}

}
\end{table*}

\subsection{Detecting Test Smells}

\gemma{} and \phimodel{} exhibit good detection performance across most test smells, correctly identifying \gemmaDetect{} and \phiDetect{} instances, respectively, when using two agents. Their performance remains consistent with one or four agents. For every test smell, at least one model achieves a perfect detection score.
\deepseek{} demonstrates the weakest performance, detecting only \deepseekDetect{} instances.
In contrast, \llama{} consistently detects all test smells across all agent configurations.

We review Agents$_{1,2}$ outputs to assess the validity of their test smell explanations.
\review{
Next, Agent$_{1}$ uses \phimodel{} to identify the \textit{Exception Handling} test smell. The other models also detect it.
\vspace{0.2em} 
\begin{mdframed}[backgroundcolor=light-gray, linecolor=black, linewidth=0.5pt]
\footnotesize
\review{
\textit{``The test method \texttt{testStackBlowOut} includes a \texttt{try-catch} block to handle exceptions manually, which is indicative of the Exception Handling test smell. The code attempts to catch an \texttt{IllegalArgumentException}, rather than using testing framework features like \texttt{assertThrows} to verify that the exception is thrown as expected.''}
}
\end{mdframed}
}
\llama{} achieved the highest detection rate, identifying all test smell instances. \gemma{} and \phimodel{} missed a few but provided more detailed and accurate explanations.
These results demonstrate that smaller models can reliably detect test smells when given concise natural language definitions. Unlike static-analysis tools, \review{models} take a semantic approach, recognizing frameworks like Mockito or JUnit without needing exhaustive method lists.

\subsection{Refactoring Test Smells}

The optimal configuration for \llama{} and \gemma{} involved using four-agents, successfully refactoring \llamaRefactor{} and \gemmaRefactor{} out of \totalExamples{} instances in one attempt, respectively.
\deepseek{} achieved its best performance with a single-agent, correctly refactoring \deepseekRefactor{} instances.
\phimodel{} performed best with two-agents, successfully refactoring \phiRefactor{} instances, demonstrating its effectiveness in a multi-agent setup.
\review{Processing \totalExamples{} smells using four-agents took \textasciitilde1h (\llama{}, \gemma{}), \textasciitilde5h (\phimodel{}), and \textasciitilde8h (\deepseek{}).}

Using single, two, or four-agents with \llama{}, \gemma{}, or \phimodel{}, correctly remove the \textit{Exception Handling} test smell. For instance, they produce the code shown in Listing~\ref{lst:example-exception-refactor}, where the \texttt{try-catch} block is removed, and the \texttt{fail} statement is replaced with an \texttt{assertThrows} statement. 
\begin{lstlisting}[basicstyle=\footnotesize, language=Java, label=lst:example-exception-refactor, caption={Correct refactoring for removing \textit{Exception Handling} using single and multi-agents with \gemma{}.}]
@Test
public void testStackBlowOut() {
    final SmallRyeConfig config = buildConfig(maps(singletonMap("foo.blowout", "${foo.blowout}")));
    assertThrows(IllegalArgumentException.class, () -> config.getValue("foo.blowout", String.class));
}
\end{lstlisting}

In some cases, using single-agent with \gemma{} to refactor results in incorrect code that alters the program's behavior by adding an extra assertion when removing the \textit{Conditional Test Logic} test smell. In contrast, using multi-agents produces a correct refactoring, effectively separating the logic into two methods. 

We can also correctly apply a refactoring to remove \textit{Duplicate Assert} using multi-agents with \phimodel{}. However, using a single-agent with \phimodel{} applies a refactoring strategy that differs from the one proposed for this smell. It incorporates a unique message for each assertion, effectively eliminating both this test smell and the \textit{Assertion Roulette} test smell.

\deepseek{} provides significantly more details about its reasoning process compared to the other models. In some cases, it initially proposes a valid refactored code or a correct refactoring strategy during its thought process but later revises its approach, ultimately leading to an incorrect solution.

For \textit{Assertion Roulette}, \phimodel{} achieved the best result, correctly detecting and refactoring 93.3\% of instances with a single-agent. In contrast, all models struggled with \textit{Conditional Test Logic}, where we evaluated only one of several possible refactoring strategies. For \textit{Duplicate Assert}, the highest success rate (70\%) was obtained by \phimodel{} using two-agents. \phimodel{} also performed best on \textit{Exception Handling}, correctly refactoring 43.3\% of cases with a single-agent. The best results for \textit{Magic Number} came from \deepseek{} with two agents, achieving 90\% accuracy.
Overall, two or four-agent setups led to better outcomes in three of the five test smells. However, for \textit{Assertion Roulette} and \textit{Exception Handling}, a single-agent was sufficient. Notably, there are 23 unique instances successfully refactored by at least one model -- \llama{} (6), \gemma{} (14), or \deepseek{} (6) -- that \phimodel{} with two-agents failed to refactor. By combining strengths across models, we achieve correct refactorings for \totalRefactor{} instances \review{(pass@1)}.

To improve performance under the default configuration, we conducted an experiment to assess the impact of temperature~\cite{temperature} settings on \phimodel{}'s ability to detect and refactor test smells. A temperature of 0.9 yielded the best results for both tasks, slightly improving performance by 3.3\%. 

\review{
\subsection{Failure Types}

The ablation study revealed distinct failure patterns across agent configurations.
\minorreview{Although adding more agents reduced certain errors, it also introduced new challenges, such as missed detections previously handled by the single-agent approach, incomplete fixes, and cases where Agent$_4$ rejected correct refactorings proposed by Agent$_3$. In the aggregated analysis, the multi-agent setup outperformed the single-agent configuration in three test smells, matched its performance in \textit{Exception Handling}, and underperformed only in \textit{Assertion Roulette}. The instance-level analysis with \phimodel{} identified 12 cases in which the single-agent succeeded but the multi-agent failed. These included missed detections in \textit{Conditional Test Logic} and \textit{Magic Number}; misplaced or invalid refactorings in \textit{Assertion Roulette}, where the model sometimes generated JUnit-4 instead of JUnit-5 messages; partial fixes in \textit{Magic Number}, where only a subset of constants was extracted or new variables were not properly initialized; and issues in \textit{Exception Handling}, where critical code was inadvertently removed along with the \texttt{try} block. In addition, one \textit{Duplicate Assert} case introduced a behavioral change.}
}

\subsection{Feedback Loop}

The feedback loop was triggered when Agents$_{2,4}$ disagreed with Agents$_{1,3}$, respectively (see Figure~\ref{fig:agentic4}).  
For detecting test smells, \gemma{} and \phimodel{} utilized the feedback loop in 1 and 9 test smell instances, respectively. In 2 instances, \phimodel{} was able to correctly identify a test smell due to the feedback loop. \llama{} did not require the feedback loop for detecting any test smells. \deepseek{} successfully detected 3 instances of the \textit{Duplicate Assert} test smell following a feedback loop. 
We employed a three-iteration feedback loop, which was sufficient for Agents$_{1,2}$ to reach a consensus on detecting the majority of test smells. For Agents$_{3,4}$ to reach a consensus on refactoring test smells, only 7.3\% of the cases required more than three iterations. By allowing additional iterations, we achieved consensus in all cases by the eighth iteration, except for a single instance of \textit{Conditional Test Logic}.

\subsection{Pull Requests}
To assess the acceptance of the refactorings generated by \phimodel{}, which achieved the best performance in our study, we submitted 10 pull requests, across five different open-source projects. 
As of this writing, \review{six} of these pull requests have been \review{reviewed and merged by human maintainers, who deemed them valuable:} two in \textit{janusgraph} (\textit{Assertion Roulette} and \textit{Exception Handling}), two in \textit{lettuce} (\textit{Magic Number} and \textit{Duplicate Assert}), and one each in \textit{opengrok} and \textit{jenkins} (both addressing \textit{Duplicate Assert}).
\review{Two} pull requests \review{were} rejected because contributors noted that the additional message in each assertion did not improve their code. 
\review{Two pull requests remain open.}

\subsection{\review{LLMs}}

\review{
In July 2025, we also run \othree{}, \claude{}, and \gemini{} using a single-agent setup on \totalExamples{} test smells. \claude{}, \othree{}, \gemini{} achieved a refactoring performance (pass@1) of \claudeRefactor{}, \othreeRefactor{}, and \geminiRefactor{}, respectively—though at significantly higher computational cost.

We allowed \phimodel{} with four-agents to make up to five attempts, rather than limiting it to just one. With pass@5~\cite{pass-k}, it successfully detected \phiDetectPassFive{} and refactored \phiRefactorPassFive{} of the test smell instances.
}
These results indicate that the open \phimodel{}, when paired with agentic workflows, can rival proprietary state-of-the-art \review{LLMs} while remaining cost-effective and executable on consumer-grade hardware.

\subsection{Prompts}

We refined the prompts iteratively. For instance, in the definition of \textit{Assertion Roulette}, the phrase ``\textit{has more than one assertion}'' proved to be more precise and effective than the commonly used ``\textit{has multiple non-documented assertion}'' found in the literature~\cite{Peruma2019Distribution}.
Early experiments overlooked prompt minimization, resulting in suboptimal outcomes -- consistent with Hsieh et al.~\cite{hsieh2024rulerwhatsrealcontext}. Providing Agent$_{4}$ with the original and refactored code plus Agent$_{3}$’s full explanation overloaded the model and reduced its effectiveness. Similarly, overly detailed smell definitions negatively impacted performance.
Although not formally tested, we observed that large prompts hindered reasoning. Streamlining definitions and instructing Agent$_{4}$ to respond concisely improved efficiency. These findings align with prior studies showing that excessive or irrelevant context degrades LLM performance~\cite{llms-irrelevant-context}.
In some cases, we found that sentence order influenced model reasoning, supporting prior work~\cite{llms-order-matters}.

\subsection{Other Languages}

\review{
We assess whether the same setup can detect and refactor test smells across different programming languages. Using a single-agent configuration, \phimodel{} successfully refactored (pass@1) 24/30, 25/30, and 29/30 test smells in Golang, JavaScript, and Python, respectively. These instances contained \textit{Assertion Roulette}, \textit{Duplicate Assert}, and \textit{Magic Number} smells in real-world projects such as \textit{kubernetes}, \textit{react-beautiful-dnd}, and \textit{django}. The model achieved performance comparable to that observed with Java test cases. 
In the JavaScript setting, it occasionally generated Java code; however, when explicitly instructed to produce JavaScript test code, it responded correctly. 
These preliminary results suggest that the proposed approach is promising for detecting and refactoring test smells across a variety of programming languages.
}

\review{
\subsection{Detecting Other Types of Test Smells}

We evaluated \phimodel{} on 30 types of test smells~\cite{lucas-sbes-nier-2024}, using real-world examples—18 in Java and 12 in other languages including Ruby, Python, JavaScript, Smalltalk, and TTCN-3.
A single-agent was used, focusing solely on detection, as the catalog does not define refactoring strategies for all smells. \phimodel{} successfully detected 28 out of 30 test smells (pass@1), providing detailed justifications for each. The two initial failures were due to ambiguous definitions, which, once clarified, enabled the accurate detection of all smells.
We also evaluated \phimodel{} on 45 instances spanning four additional test-smell types in JavaScript. It misclassified only four cases (pass@1) of \textit{OverCommented Tests}, due to an imprecise definition of when a test contains excessive comments. After clarifying the definition, it correctly detected all test smells.
}
\subsection{Larger Test Cases}

As previously noted, \review{89\% of the test cases in the 11 open-source projects analyzed contain at most 30 LOC.}
To further evaluate \phimodel{}, which demonstrated the best performance in our study, we selected 50 test smells (ten per smell type) exceeding 30 LOC from these 11 open-source projects. We employed the same four-agent setup to detect and refactor these test smells.
Although the strategy successfully identified 86\% of the test smells, it showed a decline in performance compared to previous results. The effectiveness in accurately refactoring test smells also diminished, with a \review{pass@1} rate of 28\%.

\minorreview{
\subsection{Code Smells}
As a feasibility study, we evaluated \phimodel{} on 30 Java code smells drawn from real-world projects. By making minor adjustments to the prompts—for example, replacing references to ``test'' with ``code''—\phimodel{} successfully detected all instances and correctly refactored a subset of them. These results indicate that the approach is promising for the automated detection and refactoring of code smells. We leave a deeper investigation to future work.
}

\subsection{Threats to Validity}

One potential threat to internal validity is data leakage, where test smells analyzed in this study may be part of the foundation models' training data. To mitigate this, we applied Metamorphic Testing.\review{\hyperref[fn:exemplo]{\textsuperscript{\ref{fn:exemplo}}}}
We limit our analysis to test code with a maximum of 30 LOC. While this constraint excludes longer test cases, a substantial number of test cases still fall within this criterion, ensuring a representative evaluation.
Another factor is prompt design, which can lead to generic responses instead of targeted explanations. To minimize variability, we used concise prompts across all evaluations, which were carefully reviewed by three authors of this paper.
Validating alignment between the model output and test smell definitions is also challenging. To ensure accuracy, three authors independently reviewed selected responses, verifying whether the refactored code adhered to the intended definitions.
Construct validity is limited by the small number of examples for each test smell, though they are real cases from GitHub repositories.

\section{Related Work}
\label{sec:related}

Aljedaani et al.~\cite{aljedaani2021test} compiled a comprehensive catalog of 22 test smell detection tools, the majority of which are designed for Java, SmallTalk, C++, and Scala, and 4 refactoring tools. 
Soares et al.~\cite{SoaresRGAS23} conducted a mixed-methods analysis involving 485 Java projects, exploring the adoption of JUnit-5 features to improve test code quality. 
Wang et al.~\cite{pynose} propose a tool called PyNose to detect 18 types of test smells in Python.
Lambiase et al.~\cite{lambiase2018just} introduce DARTS, an IntelliJ plugin that detects and refactors three test smells: \textit{Eager Test}, \textit{General Fixture}, and \textit{Lack of Cohesion of Test Methods}. 
Pontillo et al.~\cite{pontillo2024} proposed a machine learning-based approach to detect test smells, focusing on four specific types.  
Lucas et al.~\cite{lucas-sbes-nier-2024} investigated the capability of three LLMs to detect test smells across multiple programming languages. 

We investigate the effectiveness of agentic workflows for detecting and refactoring test smells in Java test cases. Our approach coordinates up to four-agents to automate these tasks. Results show that \phimodel{} achieves strong refactoring performance. It runs efficiently on consumer-grade hardware and perform comparably to \review{LLMs}.
A key advantage of our approach is its extensibility: new smells can be added via natural language definitions. Preliminary experiments also confirm successful application to Python, Golang\review{, and JavaScript}, highlighting its cross-language potential.
Unlike conventional methods that only detect smells or suggest code changes, our approach supports interactive conversations and explanatory feedback.

\section{Conclusions}
\label{sec:conclusions}

We evaluated \llama{}, \gemma{}, \deepseek{}, and \phimodel{} for the automated detection and refactoring of test smells using agentic workflows with one, two, and four agents. The evaluation covered \totalExamples{} instances of \totalTypes{} common test smell types from real-world projects. We assessed each model's ability to perform automated refactorings.
\phimodel{}, \gemma{}, and \llama{} successfully detected nearly all test smell instances. \phimodel{} delivered the best refactoring performance, reaching a pass@5 of \phiRefactorPassFive{}, and performed within 5\% of proprietary models like \review{\othree{}, \claude{},} and \gemini{} using a single-agent.
We submitted 10 \phimodel{}-refactored pull requests, and \review{six} were merged, showing practical utility.
Multi-agent workflows outperformed single-agent setups in three out of five test smell types. 

\section*{Acknowledgments}
We thank the reviewers for their feedback. This work was partially supported by CNPq (403719/2024-0, 310313/2022-8, 404825/2023-0, 443393/2023-0, 312195/2021-4), FAPESQ-PB (268/2025), and FAPESB.

\bibliographystyle{IEEEtran}

\begin{thebibliography}{10}
\providecommand{\url}[1]{#1}
\csname url@samestyle\endcsname
\providecommand{\newblock}{\relax}
\providecommand{\bibinfo}[2]{#2}
\providecommand{\BIBentrySTDinterwordspacing}{\spaceskip=0pt\relax}
\providecommand{\BIBentryALTinterwordstretchfactor}{4}
\providecommand{\BIBentryALTinterwordspacing}{\spaceskip=\fontdimen2\font plus
\BIBentryALTinterwordstretchfactor\fontdimen3\font minus \fontdimen4\font\relax}
\providecommand{\BIBforeignlanguage}[2]{{%
\expandafter\ifx\csname l@#1\endcsname\relax
\typeout{** WARNING: IEEEtran.bst: No hyphenation pattern has been}%
\typeout{** loaded for the language `#1'. Using the pattern for}%
\typeout{** the default language instead.}%
\else
\language=\csname l@#1\endcsname
\fi
#2}}
\providecommand{\BIBdecl}{\relax}
\BIBdecl

\bibitem{van2001refactoring}
A.~van Deursen, L.~Moonen, A.~van Den~Bergh, and G.~Kok, ``Refactoring test code,'' in \emph{XP}, 2001, pp. 92--95.

\bibitem{Peruma2019Distribution}
A.~Peruma, K.~Almalki, C.~D. Newman, M.~W. Mkaouer, A.~Ouni, and F.~Palomba, ``On the distribution of test smells in open source {Android} applications: An exploratory study,'' in \emph{CSSE}, 2019, pp. 193--202.

\bibitem{aljedaani2021test}
W.~Aljedaani et al., ``Test smell detection tools: A systematic mapping study,'' in \emph{EASE}, 2021, pp. 170--180.

\bibitem{pontillo2024}
V.~Pontillo, D.~Amoroso~d’Aragona, F.~Pecorelli, D.~Di~Nucci, F.~Ferrucci, and F.~Palomba, ``Machine learning-based test smell detection,'' \emph{Empirical Software Engineering}, vol.~29, no.~2, pp. 1--44, 2024.

\bibitem{mesaros}
G.~Meszaros, \emph{{xUnit} test patterns: Refactoring test code}. Pearson, 2007.

\bibitem{SoaresRGAS23}
E.~Soares, M.~Ribeiro, R.~Gheyi, G.~Amaral, and A.~L.~M. Santos, ``Refactoring test smells with {JUnit} 5: Why should developers keep up-to-date?'' \emph{TSE}, vol.~49, no.~3, pp. 1152--1170, 2023.

\bibitem{prompt-techniques2}
\BIBentryALTinterwordspacing
S.~Schulhoff et al., ``The prompt report: A systematic survey of prompting techniques,'' 2024. \url{https://arxiv.org/abs/2406.06608}
\BIBentrySTDinterwordspacing

\bibitem{temperature}
A.~Holtzman, J.~Buys, L.~Du, M.~Forbes, and Y.~Choi, ``The curious case of neural text degeneration,'' in \emph{ICLR}, 2020.

\bibitem{pass-k}
\BIBentryALTinterwordspacing
M.~Chen et al., ``Evaluating LLMs trained on code,'' 2021. \url{https://arxiv.org/abs/2107.03374}
\BIBentrySTDinterwordspacing

\bibitem{hsieh2024rulerwhatsrealcontext}
\BIBentryALTinterwordspacing
C.-P. Hsieh et al., ``{RULER}: {W}hat's the real context size of your long-context language models?'' 2024.  \url{https://arxiv.org/abs/2404.06654}
\BIBentrySTDinterwordspacing

\bibitem{llms-irrelevant-context}
F.~Shi et al., ``LLMs can be easily distracted by irrelevant context,'' in \emph{ICML}, vol. 202. 2023, pp. 31\,210--31\,227.

\bibitem{llms-order-matters}
X.~Chen, R.~A. Chi, X.~Wang, and D.~Zhou, ``Premise order matters in reasoning with LLMs,'' in \emph{ICML}. 2024.

\bibitem{lucas-sbes-nier-2024}
K.~Lucas, R.~Gheyi, E.~Soares, M.~Ribeiro, and I.~Machado, ``Evaluating large language models in detecting test smells,'' in \emph{SBES}, 2024, pp. 672--678.

\bibitem{pynose}
T.~Wang, Y.~Golubev, O.~Smirnov, J.~Li, T.~Bryksin, and I.~Ahmed, ``{PyNose}: {A} test smell detector for {Python},'' in \emph{ASE}. 2021, pp. 593--605.

\bibitem{lambiase2018just}
S.~Lambiase, A.~Cupito, F.~Pecorelli, A.~De~Lucia, and F.~Palomba, ``Just-in-time test smell detection and refactoring: The {DARTS} project,'' in \emph{ICPC}, 2020, pp. 441--445.

\end{thebibliography}

\end{document}